\documentclass[letterpaper, 10 pt, conference]{ieeeconf}  

\IEEEoverridecommandlockouts                              

\usepackage{graphicx}
\usepackage{algorithm}
\usepackage{fnpct}
\usepackage{footnote}
\usepackage{graphics} 
\usepackage{graphicx} 
\usepackage{epsfig} 
\usepackage{times} 
\usepackage{amsmath} 
\usepackage{amssymb}  
\usepackage{tikz}
\usepackage{dirtytalk}
\usetikzlibrary{arrows.meta, positioning, shapes}
\usepackage{url,verbatim,color,comment}
\usepackage[noend]{algpseudocode}
\usepackage{eucal}
\usepackage{amsfonts}
\usepackage{xcolor}
\usepackage{blindtext}
\usepackage{mathtools}
\usepackage{placeins}
\usepackage{fancyhdr}
\usepackage[hidelinks]{hyperref}
\hypersetup{
  colorlinks,
  citecolor= magenta,
  linkcolor=blue,
  urlcolor=blue}

\title{\LARGE \bf
The Synergy Between Optimal Transport Theory \\ and Multi-Agent Reinforcement Learning
}

\author{Ali Baheri$^1$ and Mykel J. Kochenderfer$^2$
\thanks{$^{1}$Ali Baheri is with the Department of Mechanical engineering at Rochester Institute of Technology.
        {\tt\small akbeme@rit.edu}}%
\thanks{$^{2}$Mykel J. Kochenderfer is with the Department of Aeronautics \& Astronautics at Stanford University. 
        {\tt\small mykel@stanford.edu}}}

\begin{document}

\maketitle
\thispagestyle{empty}
\pagestyle{empty}

\begin{abstract}

This paper explores the integration of optimal transport (OT) theory with multi-agent reinforcement learning (MARL). This integration uses OT to handle distributions and transportation problems to enhance the efficiency, coordination, and adaptability of MARL. There are five key areas where OT can impact MARL: (1) policy alignment, where OT's Wasserstein metric is used to align divergent agent strategies towards unified goals; (2) distributed resource management, employing OT to optimize resource allocation among agents; (3) addressing non-stationarity, using OT to adapt to dynamic environmental shifts; (4) scalable multi-agent learning, harnessing OT for decomposing large-scale learning objectives into manageable tasks; and (5) enhancing energy efficiency, applying OT principles to develop sustainable MARL systems. This paper articulates how the synergy between OT and MARL can address scalability issues, optimize resource distribution, align agent policies in cooperative environments, and ensure adaptability in dynamically changing conditions. 

\end{abstract}

\section{INTRODUCTION}

Multi-agent reinforcement learning (MARL) is a framework where multiple agents interact, learn, and adapt within shared environments \cite{marl-book}. These interactions, governed by diverse objectives and constraints, present significant challenges related to coordination, resource management, adaptability, and operational efficiency \cite{wong2023deep}. 

Optimal transport (OT) theory offers a powerful mathematical framework for comparing and transforming probability distributions in a cost-effective manner \cite{villani2009optimal}. It is concerned with finding the most efficient plan to transport mass from one distribution to another, minimizing a given cost function. OT has found applications across diverse fields due to its ability to provide geometrically and statistically meaningful ways to compare distributions \cite{peyre2019computational}. In this context, the fusion of OT with MARL represents a promising interdisciplinary strategy, designed to harness the strengths of OT in tackling the multifaceted challenges inherent in MARL.

\noindent{\textbf{Policy Alignment in MARL with OT.} We explore how OT, particularly through its Wasserstein distance metric, can be used to minimize the divergence in strategies among agents. This approach promises a more coherent and effective collaborative learning process, ensuring agents' policies are aligned towards common goals.

\noindent{\textbf{Distributed Resource Management in MARL with OT.} In scenarios where agents must efficiently share limited resources like energy or information, OT provides a principled framework to optimize this distribution. The paper delves into how OT can minimize the costs associated with resource allocation, balancing efficiency and equity among agents.

\noindent{\textbf{Addressing Non-Stationarity in MARL with OT.} The dynamic nature of MARL environments, where the ground truth shifts as agents learn and adapt, presents a significant challenge. OT's ability to adapt to changes in probability distributions is investigated as a method to manage this non-stationarity, enhancing capacity of the agents to respond to evolving environments.

\noindent{\textbf{Scalable Learning in Large-Scale MARL with OT.} As the scale of MARL systems expands, so does the complexity of managing interactions and learning. We propose using OT to decompose global learning objectives into localized tasks, facilitating scalable and efficient learning across large networks of agents.

\noindent{\textbf{Enhancing Energy Efficiency in MARL with OT.} In energy-constrained MARL scenarios, optimizing energy usage is crucial. This paper discusses how OT can be used to develop energy-efficient MARL systems.

\noindent{\textbf{Contributions.} This paper explores how the fusion of OT and MARL can lead to more efficient multi-agent systems. 

The structure of the paper is as follows: Section \ref{ref:prem} provides an overview of OT. Section \ref{ref:integration} discusses how to integrate OT principles into MARL. Section \ref{ref:limitations} examines the challenges of this integration, addressing potential computational issues.

\section{Optimal Transport}
\label{ref:prem}

OT theory, rooted in the works of Gaspard Monge and later Leonid Kantorovich, revolves around finding the most cost-effective way of transporting mass from one distribution to another \cite{villani2009optimal,kantorovich2006translocation}. Central to OT is the concept of the Wasserstein distance, also known as the Earth Mover's distance, which quantifies the \say{effort} required to transform one probability distribution into another. Originating from Monge's problem in the 18th century, which sought an optimal plan for moving soil with minimum effort, the theory has evolved, especially with Kantorovich’s formulation in the 20th century, to incorporate modern mathematical tools from linear programming and functional analysis \cite{rachev1998mass}. 

OT has found applications in diverse fields. In economics, it aids in understanding resource allocation and market dynamics \cite{galichon2021unreasonable,galichon2017survey}; in computer vision and graphics \cite{bonneel2023survey,wen2020conformal}, it helps in image retrieval and texture mixing \cite{delon2023optimal,rabin2012wasserstein}; and in machine learning \cite{balaji2020robust,flamary2016optimal,montesuma2023recent,redko2019optimal,baheri2023risk}, it is used for domain adaptation and generative modeling.

OT theory is concerned with the problem of transporting mass from one distribution to another in the most efficient manner. Formally, given two probability distributions $\mu$ and $\nu$ on spaces $X$ and $Y$, respectively, OT seeks a transport map $T: X \rightarrow Y$ that minimizes the cost of transportation. A key concept in OT is the Wasserstein distance, which is defined as the minimum cost to transport mass from one distribution to another. Mathematically, for distributions $\mu$ and $\nu$, the $p$-Wasserstein distance is given by:
\begin{equation}
W_p(\mu, \nu)=\left(\inf _{\gamma \in \Gamma(\mu, \nu)} \int_{X \times Y} d(x, y)^p d \gamma(x, y)\right)^{1 / p}
\end{equation}
where $\Gamma(\mu, \nu)$ represents all joint distributions (couplings) with marginals $\mu$ and $\nu$, and $d(x, y)$ is a distance metric on $X \times Y$.


\section{Integration of OT with MARL}
\label{ref:integration}

\subsection{OT and Policy Alignment}

Cooperative learning in MARL requires aligning the policies and objectives of individual agents towards a collective goal. This section shows how OT theory in can amplify the efficacy of cooperative learning among agents in MARL environments. In cooperative MARL, a group of agents $\left\{a_1, a_2, \ldots, a_n\right\}$ aims to learn policies $\left\{\pi_1, \pi_2, \ldots, \pi_n\right\}$ that maximize a shared reward function. The effectiveness of cooperation hinges on the alignment of these policies towards the collective goal. Each agent's policy $\pi_i$ induces a probability distribution $\mu_i$ over the state-action space $\mathcal{S} \times \mathcal{A}$. The alignment of these distributions is crucial for effective cooperation. The $p$-Wasserstein distance could be used to measure the discrepancy between these distributions. For two agents $a_i$ and $a_j$, the distance between their induced distributions $\mu_i$ and $\mu_j$ is given by:
\begin{equation}
W_p\left(\mu_i, \mu_j\right)=\left(\inf _{\gamma \in \Gamma\left(\mu_i, \mu_j\right)} \int_{\mathcal{S} \times \mathcal{A}}\left\|s-s^{\prime}\right\|^p d \gamma\left(s, s^{\prime}\right)\right)^{1 / p}
\end{equation}
where $\Gamma\left(\mu_i, \mu_j\right)$ represents the set of all joint distributions with marginals $\mu_i$ and $\mu_j$, and $\| s-$ $s^{\prime} \|$ is a metric on the state-action space. The goal is to adjust individual policies to minimize the overall Wasserstein distance between all pairs of agents' policy-induced distributions. The optimization problem is:
\begin{equation}
\min \sum_{i=1}^N \sum_{j=1, j \neq i}^N W_p\left(\mu_i, \mu_j\right)
\end{equation}
This minimization ensures that the policies of different agents are coherently aligned towards the collective objective. Implementing OT for policy alignment in MARL is conceptually promising, but choosing an appropriate metric for the state-action space is crucial and can be complex. 

The crux of this integration lies in harnessing OT’s capacity for efficiently transporting and transforming distributions to align policy distributions within MARL environments. Each agent's policy can be viewed as a probability distribution over the state-action space, reflecting their decision-making process. The Wasserstein distance could provide a means to measure the \say{effort} or \say{cost} needed to align one agent's policy distribution with another. By minimizing this distance, we can effectively guide the agents towards policy alignment.

\noindent {\textbf{Related Work.} Various approaches have been proposed to facilitate policy alignment. These include shared reward structures, where agents are incentivized to work towards common goals, and joint action learning, where agents learn policies based on the combined actions of all agents in the system \cite{chakraborty2023aligning}. Communication protocols have been explored as a means to align policies \cite{foerster2016learning}. By allowing agents to share information about their states, actions, or intended goals, these protocols can significantly improve coordination and policy alignment. A notable approach in recent years is centralized training with decentralized execution \cite{hong2022rethinking,9905866}. Here, agents are trained in a centralized manner, allowing them to learn about each other's policies, but execute their learned policies independently. Policy alignment in MARL has found applications in various domains such as autonomous vehicle coordination and robotics \cite{schmidt2022introduction}.

\subsection{OT and Distributed Resource Management}

In leveraging OT for distributed resource management in MARL, we focus on the transportation of resources as a probabilistic measure. This approach aligns with the core principles of OT, where the Wasserstein distance is used to quantify the efficiency of transporting one distribution to another.

In a MARL system with resources $R=\left\{r_1, r_2, \ldots, r_m\right\}$ and agents $A=\left\{a_1, a_2, \ldots, a_n\right\}$, consider each resource $r_j$ as a mass that needs to be distributed among agents. Each agent $a_i$ has a requirement or demand distribution $\mu_i$, and the total available resources form a supply distribution $\nu$. The transportation of resources from the total supply $\nu$ to meet the demands $\mu_i$ of each agent can be modeled using the Wasserstein distance. The goal is to find an OT plan that minimizes the Wasserstein distance, hence the transportation cost, between the supply and demand distributions. The mathematical formulation becomes:
\begin{equation}
W_p\left(\nu, \mu_i\right)=\left(\inf _{T \in \Gamma\left(\nu, \mu_i\right)} \int_{\mathcal{R} \times \mathcal{A}}\|r-a\|^p d T(r, a)\right)^{1 / p}
\end{equation}
Here, $\Gamma\left(\nu, \mu_i\right)$ represents the set of all possible transport plans (joint distributions) with marginals $\nu$ and $\mu_i$, and $\|r-a\|$ is a metric representing the cost of transporting resources. The OT problem in this context aims to minimize the total transportation cost across all agents, ensuring that resources are allocated efficiently and in accordance with agents' demands:
\begin{equation}
\min \sum_{i=1}^n W_p\left(\nu, \mu_i\right)
\end{equation}
This formulation respects the constraints of resource availability and agent requirements, optimizing the overall distribution of resources in the MARL system.

The point of this integration lies in using OT’s mathematical framework to coordinate the allocation of resources among agents. By defining a tailored cost function within the OT model, we can effectively capture and minimize the costs associated with resource distribution, such as energy expenditure, transit time, or spatial constraints. This approach could augment the overall efficiency of resource allocation and introduces a level of fairness and adaptability that is often unattainable with traditional MARL algorithms. The dynamic nature of MARL systems, characterized by fluctuating resource demands and environmental states, further highlights the suitability of OT. Its inherent flexibility to \emph{recalibrate} transport plans in response to changing conditions ensures that the resource distribution remains optimal over time. 

\noindent{\textbf{Related Work.}} 
Several works have proposed various strategies for resource allocation in MARL. These include auction-based mechanisms, where agents bid for resources \cite{talebiyan2023auctions,braquet2021greedy}, and cooperative strategies, where agents share resources based on collective goals \cite{li2021multi}. There is an open discussion in current research regarding the balance between decentralized versus centralized approaches for resource allocation \cite{lu2022centralized}. Recent studies have explored the integration of communication protocols with resource management in MARL. Distributed resource management in MARL has been applied in various domains, such as networked robotics, traffic control, and energy grids \cite{10123921,gielis2022critical}.

\subsection{OT and Non-Stationarity in MARL}

Non-stationarity, characterized by the changing dynamics in a multi-agent environment, poses a significant challenge in MARL. We propose OT to address non-stationarity in MARL systems. In MARL, agents continually adapt their policies based on the evolving environment, leading to non-stationary dynamics. This non-stationarity complicates learning because the ground truth each agent tries to learn keeps shifting. OT could offer a powerful framework to adapt to changes in probability distributions, which, in the context of MARL, correspond to the evolving strategies and states of agents. We discuss how the Wasserstein metric provides a natural way to quantify the shifts in distributions representing policies or environment states over time. OT can be used to model the non-stationarity in MARL as a transportation problem, where the goal is to find an optimal plan that minimizes the cost of adapting to the evolving environment. We propose to explore algorithms that use the Wasserstein distance to quantify the rate of change in the environment and adjust the learning strategies. Using OT, the learning rate of agents can be dynamically adjusted, becoming more responsive during periods of rapid environmental change and more stable during periods of relative constancy. 

Consider MARL with $N$ agents, where each agent $i$ has a policy $\pi_i$ that generates a distribution over states $\mu_i$ at any given time $t$. The state space is denoted by $\mathcal{S}$. For agent $i$, the shift in its state distribution between two consecutive time steps $t$ and $t+1$ can be quantified using the $p$-Wasserstein distance:
\begin{equation}
W_p\left(\mu_t^i, \mu_{t+1}^i\right)=\left(\inf _{\gamma_t^i \in \Gamma\left(\mu_t^i, \mu_{t+1}^i\right)} \int_{\mathcal{S} \times \mathcal{S}}\left\|s-s^{\prime}\right\|^p d \gamma_t^i\left(s, s^{\prime}\right)\right)^{1 / p}
\end{equation}
where $\gamma_t^i$ is an optimal transport plan between distributions at times $t$ and $t+1$, and $\| s-$ $s^{\prime} \|$ is a distance metric in the state space $\mathcal{S}$. The learning rate $\alpha_t^i$ for policy updates of agent $i$ at time $t$ can be modulated based on the Wasserstein distance, reflecting the extent of environmental changes.
\begin{equation}
\alpha_t^i=f\left(W_p\left(\mu_t^i, \mu_{t+1}^i\right)\right)
\end{equation}
where $f$ could be a function mapping the Wasserstein distance to a learning rate, ensuring higher responsiveness during periods of rapid change. One could incorporate the adaptive learning rate into the policy update mechanism for each agent. This could be integrated into gradient-based learning algorithms, where the policy gradient is scaled by $\alpha_t^i$.
\begin{equation}
\pi_{t+1}^i=\pi_t^i+\alpha_t^i \nabla_{\pi_t^i} J\left(\pi_t^i\right)
\end{equation}
Here, $J\left(\pi_t^i\right)$ represents the objective function for agent $i$, typically related to expected returns.

\noindent{\textbf{Related Work.}} Research in MARL recognized non-stationarity as a fundamental problem \cite{hernandez2018multiagent}. These studies noted that the independent learning approach of agents leads to a non-stationary environment, affecting convergence and stability. A subset of work has focused on modeling the behaviors of other agents to mitigate non-stationarity. In that line, \cite{lowe2017multi} introduced models that explicitly consider the policies of other agents, thereby adapting to their strategies. The paradigm of centralized training with decentralized execution addresses non-stationarity by allowing agents to learn about each other during the training phase \cite{foerster2016learning}.

\subsection{OT and Scalable Multi-Agent Learning}

As the number of agents increases, the complexity of interactions and the computational burden escalate. We propose that OT can be used to address these scalability issues. In large MARL systems, the global learning objective can be decomposed into smaller, localized objectives using OT principles. This involves breaking down the global state and action spaces into subsets, each managed by a group of agents. The OT framework could be used to optimally assign and balance these subsets, ensuring that the localized objectives align with the global goal. Each agent group employs a localized version of the Wasserstein metric to optimize its policy within its subset of the state-action space. This localized learning reduces the computational burden while maintaining an overall coherence with the global objective, as the Wasserstein distance ensures a consistent measure of divergence across all subsets. 

In large MARL systems, task allocation becomes a critical challenge. Here, OT could be used to optimally distribute tasks among agents, ensuring that each agent operates at its maximum efficiency without being overburdened. This distribution considers both the capabilities of agents and the current demand of the system. In extremely large systems, a hierarchical decomposition approach could be employed. This involves creating multiple layers of agent groups, each with its own localized OT-based optimization problem. The solutions at each layer feed into the next, maintaining alignment with the global objective. Furthermore, inter-layer coordination between different layers could be managed through an OT framework, which optimizes the flow of policies and rewards up and down the hierarchy. This would ensure that local optimizations at lower layers contribute effectively to the global objectives at higher layers.

\noindent{\textbf{Related Work.}} To address scalability, many studies have focused on decentralized learning approaches \cite{kao2022decentralized,li2020cooperative}. These allow agents to make decisions based on local information, reducing the computational burden. Hierarchical learning structures are another solution for scalability. By organizing agents into hierarchies or modules, the complexity can be handled more effectively. Recent advancements have explored the use of networked and graph-based approaches for scalable MARL \cite{du2022scalable,Gu2021MeanFieldMR}. Transfer learning and multi-task learning techniques have been proposed to enhance scalability \cite{mai2023deep,wang2022experience,9535269,liang2022continuous}. Scalable MARL has applications in large-scale systems like traffic management \cite{luan2022marl,9295660}, smart grids \cite{gavriluta2020cyber,mahela2020comprehensive}, and distributed control systems \cite{charbonnier2022scalable}. 

\subsection{OT and Energy Efficiency}

MARL systems, particularly in large-scale and complex environments, face the challenge of high energy consumption due to the computational demands of continuous learning and decision-making processes. We can develop variants of the Wasserstein metric that incorporates energy consumption as a weight. This metric quantifies not just the efficiency of task distribution among agents but also the energy cost associated with each task. By minimizing this energy-weighted Wasserstein distance, the MARL system can dynamically allocate tasks in a way that optimizes for both performance and energy efficiency.

We could also design algorithms that continually adapt the \say{transport paths} of computational tasks and information flow among agents. These paths are optimized based on real-time energy consumption data, dynamically rerouting tasks to agents with lower energy constraints or to times of day when energy is more abundantly available (e.g., off-peak hours in grid-connected systems). For environments with strict energy limitations, such as remote sensors or space exploration robots, the algorithms prioritize essential tasks and dynamically allocate resources to maximize operational time while maintaining essential functions. In systems equipped with energy-harvesting capabilities, OT algorithms can be used to optimally allocate and store harvested energy, prioritizing tasks based on their urgency and energy requirements.

\noindent{\textbf{Related Work.}} Studies have focused on optimizing the computational and operational aspects of learning processes, reducing the energy consumption of agents during training and execution \cite{ye2023toward}. Research in MARL under resource constraints include energy-limited settings such as wireless sensor networks or mobile robotics \cite{sahraoui2022schedule}. Papers in this domain explore strategies for agents to maximize their performance while minimizing energy usage, often involving trade-offs between task completion and power consumption \cite{10195952}. Several studies have incorporated energy metrics into the reward function of MARL algorithms \cite{9919871,qi2016energy}. Some have studies collaborative strategies in MARL, where agents work together to achieve energy efficiency \cite{su2021energy}. This includes cooperative approaches for sharing resources like battery power or computational capacity to extend the operational life of the system \cite{10339164}. Application-specific research in domains like autonomous vehicle fleets \cite{hua2022multi}, smart grids \cite{9764664}, and IoT systems \cite{li2022applications} has highlighted the importance of energy-efficient MARL. In these applications, the goal is often to optimize system-wide energy usage while ensuring the effective performance of each agent.

\section{Challenges}
\label{ref:limitations}

There are several challenges in integrating OT with MARL. A primary obstacle is the computational complexity inherent in OT, particularly when applied to the high-dimensional and dynamic spaces characteristic of MARL. The computational burden of calculating the Wasserstein metric scales significantly with the size of state and action spaces, posing challenges in real-time applications \cite{fan2022complexity}. Furthermore, the dynamic nature of MARL systems necessitates continuous recalibration of OT calculations, exacerbating computational demands. To address these computational challenges, one potential solution is the development of approximation algorithms for the Wasserstein distance. Leveraging techniques such as entropic regularization can provide a more computationally tractable approximation of the Wasserstein metric \cite{clason2021entropic}. Additionally, employing machine learning models, tailored to approximate OT calculations, can accelerate the process, making it more feasible for dynamic MARL environments \cite{montesuma2023recent}.

Another critical challenge is scalability, particularly in handling large numbers of agents. The complexity of computing OT metrics across large networks of agents can offset the benefits of optimal distribution and task allocation. Here, decentralized and hierarchical approaches to OT computation could present a viable solution \cite{lee2019hierarchical}. By decomposing the global OT problem into smaller, localized sub-problems, and solving them within agent clusters, the system can achieve scalability while maintaining the integrity of the OT framework. 

\section{Conclusions}

This paper explored integrating OT with MARL, addressing some of the critical challenges in the field. We delved into how OT's strengths in handling distributions and minimizing transportation costs can enhance MARL in areas such as policy alignment, resource management, adaptability to non-stationary environments, scalability, and energy efficiency. This integration, while showcasing promising potential, also highlights areas for further research and development. Future work can focus on refining the computational efficiency of implementing OT in large-scale MARL systems and exploring real-world applications in greater depth. By continuing to explore this integration, we aim to make MARL systems more adaptable, efficient, and capable of handling complex real-world tasks.

\bibliographystyle{IEEEtran}
\bibliography{aamas}

\end{document}